\documentclass[twocolumn,aps,prl,superscriptaddress,longbibliography]{revtex4-1}
\usepackage[colorlinks,bookmarks=false,citecolor=blue,linkcolor=red,urlcolor=blue]{hyperref}
\usepackage{verbatim}
\usepackage{color}
\usepackage{amsmath,amssymb,bm,braket,float,mathtools}
\usepackage{graphicx,xfrac,appendix}
\usepackage[english]{babel}
\makeatletter

\usepackage{epstopdf}

\makeatother
\begin{document}
\title{Wigner distribution, Wigner entropy and Quantum Refrigerator of a One-Dimensional Off-diagonal Quasicrystal }

\author{Shan Suo}
\thanks{The authors contribute equally to the work. }
\affiliation{Xingzhi College, Zhejiang Normal University, Lanxi 321100, China}
\author{Ao Zhou}
\thanks{The authors contribute equally to the work. }
\affiliation{Department of Physics, Zhejiang Normal University, Jinhua 321004, China}

\author{Yanting Chen}
\affiliation{Xingzhi College, Zhejiang Normal University, Lanxi 321100, China}

\author{Shujie Cheng}
\thanks{chengsj@zjnu.edu.cn}
\affiliation{Xingzhi College, Zhejiang Normal University, Lanxi 321100, China}
\affiliation{Department of Physics, Zhejiang Normal University, Jinhua 321004, China}

\author{Gao Xianlong}
\thanks{gaoxl@zjnu.edu.cn}
\affiliation{Department of Physics, Zhejiang Normal University, Jinhua 321004, China}

\begin{abstract}
We investigate an off-diagonal quasicrystal featuring simultaneous off-diagonal and diagonal quasiperiodic modulations.
By analyzing the fractal dimension, we map out the delocalization-localization phase diagram. We demonstrate that delocalized and localized states can be distinguished via the Wigner distribution, while extended, critical, and localized phases are separated using the Wigner entropy.
Furthermore, we explore the quantum thermodynamic properties, revealing that localized states facilitate the emergence of a quantum heater mode,
alongside the appearance of a refrigerator mode. These findings enhance our understanding of localization phenomena and expand the thermodynamic
applications of quasiperiodic systems.
\end{abstract}
\maketitle

\section{Introduction}
Anderson localization constitutes a fundamental quantum effect in condensed-matter physics,
shaping our comprehension of electron dynamics within disordered or quasiperiodic structures \cite{ref1}.
According to scaling theory, three-dimensional systems exhibit a metal-insulator transition
that separates an extended phase-allowing free electron propagation-from a localized regime
where electronic motion is spatially restricted \cite{ref2,ref4_1}. Investigations into this phenomenon are
essential for deciphering how particles behave in complex disordered environments and for
guiding the design and measurement approaches in optical lattice experiments \cite{ref3,ref4,ref5}. Experimental
confirmations of Anderson localization have been achieved across multiple physical setups.
For example, ultracold atomic gases offer tunable disorder through precise manipulation of
interactions and potentials, enabling direct observation of the transition between delocalized
and localized states \cite{ref5,ref6,ref7,ref8,ref9,ref10,ref11,ref12,ref12_1,ref12_2,ref12_3}. Similarly, photonic quasicrystals demonstrate localization behavior,
with such studies aiding the understanding of light propagation and supporting the creation
of advanced optical devices \cite{ref13,ref14,ref15,ref16,ref17,ref18,ref19,ref19_1}.

Depending on the underlying physical mechanisms, Anderson localization manifests in different ways.
In quasiperiodic systems, once the quasiperiodic potential surpasses a critical threshold,
all quantum states become spatially confined \cite{ref2}. For quasiperiodic models
incorporating short- or long-range hoppings \cite{ref21,ref22,ref23,ref25,ref26_1,ref26_2}, shallow modulations \cite{ref29}, as well as
generalized quasiperiodic modulations \cite{ref27,ref28,ref29,ref30,ref31,ref32,ref33,ref34,ref35,ref35_4,ref35_6,ref35_7}, localization arises only at certain energy levels;
these are separated from delocalized states by mobility edges. This results in an intermediate phase
that is neither entirely extended nor fully localized. Recent studies have revealed a hidden self-duality property
in certain quasiperiodically modulated systems, deepening theoretical insight into Anderson
localization and the formation of mobility edges \cite{ref35_8,ref35_9,ref35_10}.

Despite the established use of inverse participation ratios and fractal dimensions as tools to distinguish quantum phases in quasiperiodic systems, the Wigner distribution \cite{ref46,hilleryDistributionFunctionsPhysics1984,kenfackNegativityWignerFunction2004,banerjiEntanglementMeasureUsing2014,simonPeresHorodeckiSeparabilityCriterion2000,taghiabadiRevealingQuantumCorrelation2016,siyouriNegativityWignerFunction2016} has emerged as a powerful phase-space framework to differentiate delocalized and localized quantum states \cite{luWignerDistributionWigner2025}. This approach leverages the phase-space representation of quantum states, encoding both position and momentum correlations, which are inaccessible to traditional real-space or momentum-space analyses. The Wigner distribution’s inherent ability to capture quantum interference effects-manifested through negative regions in its phase-space density-provides a unique fingerprint for non-classical states, such as those exhibiting Anderson localization or critical behavior.
Building on this foundation, the concept of Wigner entropy, defined as the Shannon entropy of the Wigner distribution’s magnitude \cite{vanherstraetenQuantumWignerEntropy2021} has been developed to quantify the complexity of mobility edges in phase space   \cite{luWignerDistributionWigner2025}.
These advances motivate further inquiry into whether Wigner entropy could also help
classify different quantum phases, such as extended, critical, and localized phases.
Beyond fundamental localization properties, research into quasiperiodic structures
has extended to functional applications. Systems featuring mobility edges
and intermediate phases, for instance, have been proposed for engineering energy
current rectification \cite{ref36,ref36_2,ref36_3} and designing superradiant light sources \cite{ref37}.
Recent work further shows that the extended phase in such systems can support the
construction of quantum heat engines \cite{ref40_1}, whereas the critical phase appears
more suitable for implementing quantum heater \cite{ref40_1}. Both extended and critical
phases also offer advantages for realizing quantum accelerator \cite{ref40_1}.
This raises an intriguing question: what role might the localized phase play in the
development of quantum thermodynamic devices? In addition, quantum thermal cycles
employing quasiperiodic systems as the working medium, particularly in the context of quantum refrigerator modes,
remain unexplored. If a quasiperiodic system encompassing extended, critical, and localized phases
serves as the working medium for a quantum Otto thermal cycle, could a refrigerator mode emerge?

This work is organized as follows. Section \ref{S2} studies the off-diagonal
quasicrystal with both diagonal and off-diagonal modulations and gives out the localization phase
diagram by the fractal dimension. Section \ref{S3} studies Wigner distributions
and the Wigner entropy. Section \ref{S4} studies the thermodynamical applications
of the quantum Otto cycle with the off-diagonal quasicrystal being the working medium.
A summary is presented in Sec.~\ref{S5}.

\begin{figure}[htp]
		\centering
		\includegraphics[width=0.5\textwidth]{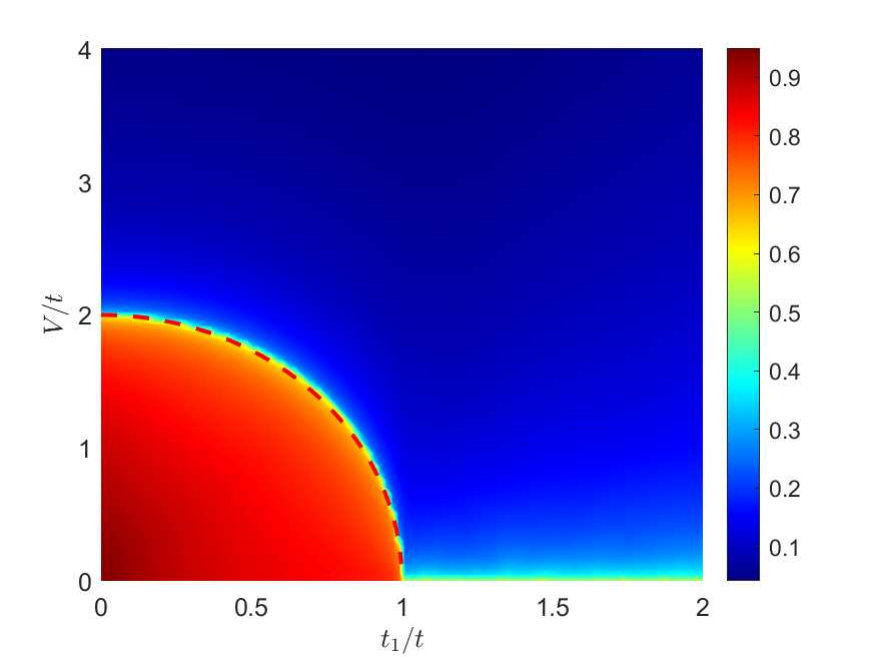}\\
		\caption{(Color Online) The phase diagram shows how fraction dimension $D$
changes with the ratios $V/t$ and $t_{1}/t$. The color corresponds to the numerical values
of $D$. The red dashed line denotes the phase boundary, which serves to separates the
extended phase ($D \sim 1$) from the localized ($D \sim 0$) and
critical phase ($D \sim 0.5$).  The involved parameter is $L=2584$.
}\label{f1}
\end{figure}

\section{Model and localization phase diagram}\label{S2}
An off-diagonal quasicrystal with both off-diagonal and diagonal quasiperiodic modulations is studied,
and its Hamiltonian is given by
\begin{equation}
H=\sum_{n}\left(t_{n}\hat{c}^{\dag}_{n}\hat{c}_{n+1}+H.c.\right)+\sum_{n}V_{n}\hat{c}^{\dag}_{n}\hat{c}_{n},
\end{equation}
where $t_{n}=t+t_{1}\cos(2\pi\alpha n)$ and $V_{n}=V\cos(2\pi\alpha n)$. $t$ is the
unit of energy. $t_{1}$ and $V$ are tunable hopping parameter and potential amplitude, respectively.
$\alpha=\frac{\sqrt{5}-1}{2}$ is the incommensurate number which leads to the term `quasiperiodic modulation'.

\begin{figure}[htp]
		\centering
		\includegraphics[width=0.5\textwidth]{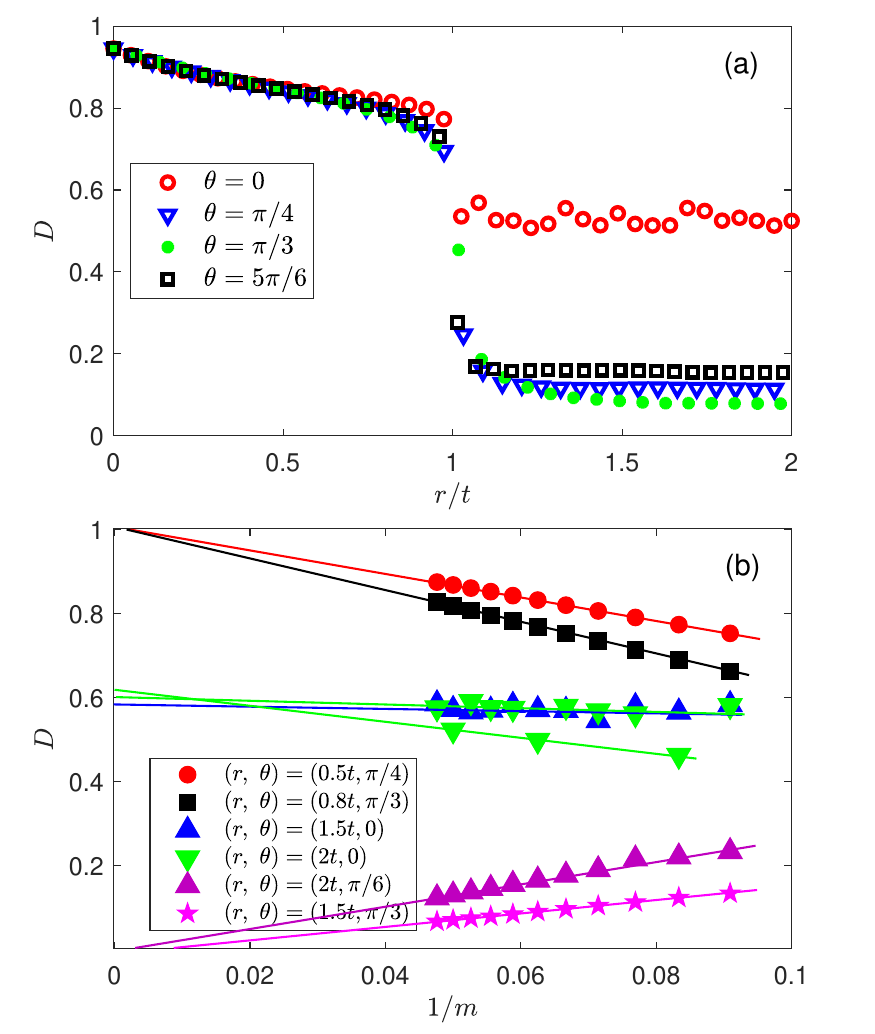}\\
		\caption{(Color Online)(a) Fractal dimension $D$ as functions of $r$ under
$\theta=0$, $\pi/4$, $\pi/3$, and $5\pi/6$.
(b) Finite size analysis of $D$ under different $(r,\theta)$, with $1/m$ the inverse Fibonacci index.
}\label{f2}
\end{figure}

The inverse participation ratio (IPR) is a commonly used tool for determining the localization
and delocalization properties of wave functions of a system.
For a normalized wave function $\psi_{j}=\sum^{L}_{n=1}\phi_{j}(n)\hat{c}^{\dag}_{n}|0\rangle$, where
$j$ is the index of the energy level which is arranged in ascending order,
the associated ${\rm IPR}_{j}$ is given by
\begin{equation}
{\rm IPR}_{j}=\sum^{L}_{n=1}|\phi_{j}(n)|^{4}.
\end{equation}
For the extended states, critical states, and localized states, their corresponding
${\rm IPR}$ respectively have the following characteristics: ${\rm IPR} \rightarrow \frac{1}{L}$
(which tends to zero under large system size),
$0<{\rm IPR}<1$, and ${\rm IPR} \rightarrow 1$ \cite{ref20,ref23}.
By averaging the ${\rm IPR}$ of all energy levels, we can obtain the
mean inverse participation ratio (MIPR), which is defined as
${\rm MIPR}=\sum^{L}_{j=1}{\rm IPR}_{j}/L$. With ${\rm MIPR}$,
we can distinguish among extended phase, critical phase and localized phase.
In the following, we will employ the scaling index of ${\rm MIPR}$, i.e.,
the fractal dimension\cite{liuFateTopologicalStates2017,wangPhaseDiagramNonAbelian2016,liuPhaseDiagramGeneralized2018} $D$ to characterize the potential quantum phases of the
system, and it is defined as $D=-\frac{\ln {\rm MIPR}}{\ln L}$. According to
the characteristics of inverse participation ratio, the corresponding MIPR
of extended phase also scales as $1/L$, so the corresponding fractal dimension
$D$ tends to $1$. The ${\rm MIPR}$ of the critical phase is still within the
interval $(0,1)$, so the $D$ corresponding to the critical phase is
within $(0,1)$. The ${\rm MIPR}$ corresponding to the localized phase
tends to $1$, so the $D$ corresponding to the localized phase tends to $0$.

Choosing the system size $L=2584$, the phase diagram illustrating the fractal
dimension $D$ in $V$-$t_{1}$ parameter space is plotted in Fig.~\ref{f1},
in which the color represents the fractal dimension $D$.
In fact, the earlier studies have established that the
system undergoes a phase transition between the extended phase
and the critical phase for the $V=0$ case \cite{ref43}.
It can be observed from Fig.~\ref{f1}(a) that at $V=0$, the fractal dimension
$D$ corresponding to the extended phase approaches $1$, whereas
the $D$ value associated with the critical phase converges to $0.5$.
Drawing on this finding, we may deduce that under the condition of
finite $V$, the quarter-elliptic region where $D$ is close to $1$ corresponds
to the extended phase. In contrast, the area outside this region exhibits
a $D$ value that tends toward $0$, which indicates that it belongs to the
localized phase.

In contrast to the non-Hermitian case \cite{Wen_Chen}, the restoration
of the system's Hermiticity enlarges the region of the extended phase.
To identify the phase boundary of the extended-localized transition,
we introduce the transformation $r=\sqrt{\frac{V^{2}}{4}+t^{2}_{1}}$
where $\theta={\rm atan}(V/t_{1}) \in \left[0,2\pi\right]$. For the system
with $L=1597$, the plots of $D$ against $r$ under various $\theta$ values are
presented in Fig.~\ref{f2}(a). It is intuitively observable that for all
distinct $\theta$, the $D$-$r$ curves that characterize the extended-localized (critical)
transition all exhibit an abrupt jump at the critical value $r_{c}/t=1$. This
critical condition corresponds to the phase boundary $\sqrt{\frac{V^{2}}{4}+t^{2}_{1}}=t$,
which is depicted as the red dashed line in Fig.~\ref{f1}.

To further verify the above mentioned conclusions we have drawn,
we perform the finite size analysis on $D$. We consider a system
where the system size $L$ equals the $m$-th Fibonacci number $F_{m}$
and the incommensurate parameter $\alpha$ is replaced by $\alpha=F_{m-1}/F_{m}$.
In the limit of $1/m \rightarrow 0$ (extrapolation limit),
 $D \rightarrow 0$ indicates a localized state,
$0< D< 1$ points to a critical state, and $D \rightarrow 1$ signifies
an extended state. We choose representative parameter points in various phases to compute $D$.
As depicted in Fig.~\ref{f2}(b), we observe that the relevant $D$ approaches
$1$ at the parameter sites $\left(r,\theta\right)=\left(0.5t,\frac{\pi}{4}\right)$
and $\left(r,\theta\right)=\left(0.8t,\frac{\pi}{3}\right)$. This observations
confirm that the system resides in the extended phase under these conditions.
As anticipated, the corresponding $D$ values fall within the
interval $\left(0,1\right)$ in the thermodynamic limit when
$\left(r,\theta\right)=\left(1.5t,0\right)$ and $\left(r,\theta\right)=\left(2t,0\right)$.
These results clearly reveal the system's critical-phase characteristics in such
parameter regimes. When $\left(r,\theta\right)=\left(1.5t,\frac{\pi}{3}\right)$ and
$\left(r,\theta\right)=\left(2t,\frac{\pi}{6}\right)$, the associated $D$
extrapolates to $0$. This outcome identifies that the system is in the localized
phase at these parameter points.

To further validate the conclusions we have derived above, we conduct
finite-size analysis on the parameter $D$. Specifically, we consider a system
where the system size $L$ is set equal to the $m$-th Fibonacci number $F_{m}$,
and the incommensurate parameter $\alpha$ is substituted with $\alpha = F_{m-1}/F_{m}$.
In the extrapolation limit where $1/m \to 0$, a convergence of $D$ to $0$ denotes a
localized state, a range of $0 < D < 1$ indicates a critical state, and an
approach of $D$ to $1$ signifies an extended state. We select representative
parameter points across different phases to calculate $D$. As illustrated in
Fig.~\ref{f2}(b), the corresponding $D$ values are found to approach $1$ at the
parameter coordinates $\left(r,\theta\right)=\left(0.5t,\frac{\pi}{4}\right)$ and
$\left(r,\theta\right)=\left(0.8t,\frac{\pi}{3}\right)$. These observations confirm
that the system exists in the extended phase under these parameter conditions.
Consistent with expectations, when $\left(r,\theta\right)=\left(1.5t,0\right)$ and
$\left(r,\theta\right)=\left(2t,0\right)$, the relevant $D$ values fall within the
interval $\left(0,1\right)$ in the thermodynamic limit. These results explicitly
demonstrate the system's critical-phase properties in such parameter regimes.
For the parameter points $\left(r,\theta\right)=\left(1.5t,\frac{\pi}{3}\right)$
and $\left(r,\theta\right)=\left(2t,\frac{\pi}{6}\right)$, the associated $D$ values
extrapolate to $0$, which identifies that the system is in the localized phase at these sites.

\section{Wigner Distribution and Wigner entropy}\label{S3}
A recent investigation \cite{luWignerDistributionWigner2025} revealed that the quantum phase space
representation, specifically the Wigner distribution, is capable of distinguishing
among the extended state, the critical state, and the localized state. Building on the Wigner
distribution, the Wigner entropy can be further derived; subsequent findings indicate
that the Wigner entropy of critical states is the largest, that of extended states falls
in the middle range, and that of localized states is the smallest. Leveraging this
characteristic, the mobility edge can also be identified. These findings offer
valuable insights for our subsequent research into whether the Wigner entropy can be
employed to discriminate among the extended phase, the critical phase, and the localized phase.

For a given wave function $\ket{\psi}$, the Wigner distribution function $W(x,p)$ can be
derived via the integral expression presented below \cite{ref46,hilleryDistributionFunctionsPhysics1984,kenfackNegativityWignerFunction2004,banerjiEntanglementMeasureUsing2014,simonPeresHorodeckiSeparabilityCriterion2000,taghiabadiRevealingQuantumCorrelation2016,siyouriNegativityWignerFunction2016}:
\begin{equation}
 W(x,p)=\frac{1}{2\pi\hbar}\int^{\infty}_{-\infty}\bra{x-\frac{y}{2}}\hat{\rho}\ket{x+\frac{y}{2}}e^{-\frac{ipy}{\hbar}}dy,
\end{equation}
where $x$ and $p$ represent the coordinate and momentum in phase space, respectively;
$\hbar$ denotes the reduced Planck constant, and $\hat{\rho}=|\psi\rangle\langle\psi|$ is the
density matrix. Building on the Wigner distribution, the Wigner entropy (denoted as $W_{S}$)
can be further derived \cite{vanherstraetenQuantumWignerEntropy2021}. Taking into account the negativity of $W(x,p)$,
the final calculation of $W_{S}$ adopts the following definition:
\begin{equation}
W_{S}=-\iint W(x,p) \ln{|W(x,p)|}dxdp,
\end{equation}
where the integral range for $x$ is $\left[-L/2,L/2\right]$, and the corresponding integral interval
for $p$ is $\left[-\pi,\pi\right]$.

\begin{figure}[htp]
  \centering
  \includegraphics[width=0.5\textwidth]{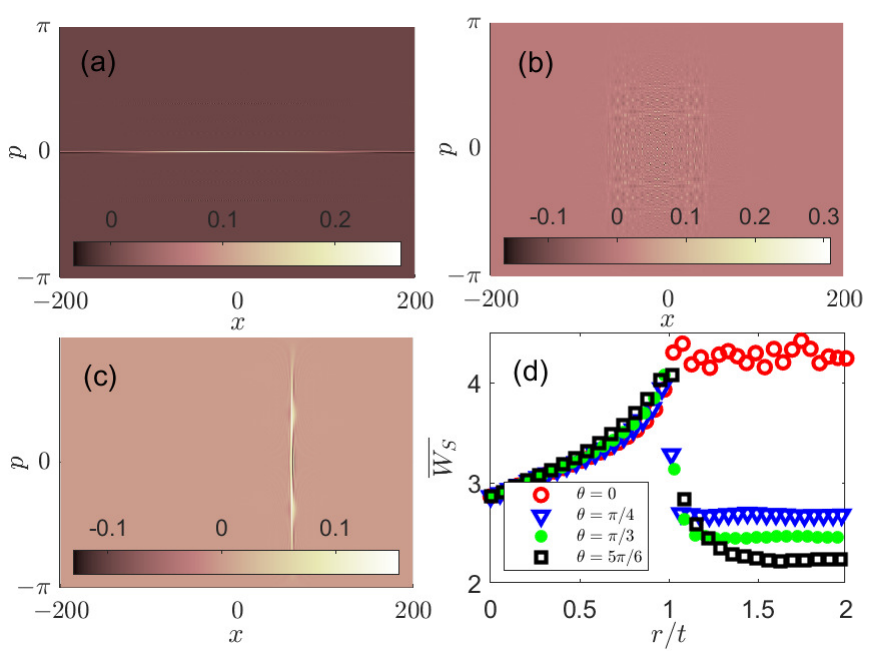}\\
  \caption{(Color Online)
  (a) Wigner distribution $W(x,p)$ of an extended state under $r=0.5t$ and $\theta=\frac{\pi}{4}$;
  (b) Wigner distribution $W(x,p)$ of a critical state under $r=1.5t$ and $\theta=0$;
  (c) Wigner distribution $W(x,p)$ of a localized state under $r=2t$ and $\theta=\frac{\pi}{6}$;
  (d) The mean Wigner entropy $\overline{W_{S}}$ as a function of $r$ under various $\theta$.
  Other involved parameter is $L=400$.
  }
  \label{f3_0}
\end{figure}

With the system size set to $L=400$, the Wigner distribution $W(x,p)$ of a typical
extended state is depicted in Fig.~\ref{f3_0}(a), that of a typical critical
state is presented in Fig.~\ref{f3_0}(b), and the Wigner distribution
corresponding to a typical localized state is plotted in Fig.~\ref{f3_0}(c).
Here, the typical states are the ground states. It can be observed that the
$W(x,p)$ of the extended state remains extended along the $x$ branch, while
it is relatively concentrated in the $p$ direction, primarily localized
around $p=0$. For the critical state, the $W(x,p)$ exists on both the $x$ branch
and $p$ branch, exhibiting a certain level of broadening. In contrast,
the $W(x,p)$ of the localized state remains localized along the $x$ branch
yet appears extended along the $p$ branch. For a wave function, we can obtain
the Wigner entropy corresponding to different wave functions by statistically
analyzing the Wigner distribution, thereby distinguishing different wave functions.
It should be noted that for a quantum phase, the characteristics of the wave functions
corresponding to all energy levels and the Wigner distribution are similar. Thus,
we can utilize the physical quantity that characterizes the common features of the
wave function and the Wigner distribution, namely the mean Wigner entropy $\overline{W_{S}}$,
i.e., $\overline{W_{S}}=\sum^{L}_{j=1}W^{j}_{s}$, to represent different quantum phases,
thereby characterizing the Anderson localization phase transition. Considering different
$\theta$ values, the variation curves of the mean Wigner entropy $\overline{W_{S}}$ with
respect to $r$ are shown in Fig.~\ref{f3_0}(d). It can be observed that the
$\overline{W_{S}}$ undergoes a jump at $r=t$. The $\overline{W_{S}}$ of the critical
phase is the largest, followed by the extended phase, and the localized phase is
the smallest. These findings support the fact that Wigner entropy can not only
distinguish wave functions with different properties, but also quantum phases with
different properties.

\section{Thermodynamical applications}\label{S4}

Recently, it has been revealed that quasiperiodic systems have rich thermodynamic applications, 
such as quantum heat engine, quantum heater and quantum accelerator.
utilities in fields such as quantum heat engine, quantum heater, and quantum accelerator.
In particular, it has been proven that the extended phase serves to sustain the working
mode of quantum heat engine, while the critical phase is beneficial for maintaining the
working modes of quantum heater \cite{ref40_1}. In the present section, we
focus on exploring the thermodynamic applications of this generalized Aubry-Andr\'{e} (AA)
model incorporating both diagonal and off-diagonal quasiperiodic modulations. Our core
objectives are to verify whether the extended phase of this model still contributes to
preserving the working mode of quantum heat engine, and to clarify which specific
working mode the localized phase tends to favor, as well as to explore whether there
exists a fourth thermodynamic working mode, namely the quantum refrigerator.

\begin{figure}[htp]
		\centering
		\includegraphics[width=0.5\textwidth]{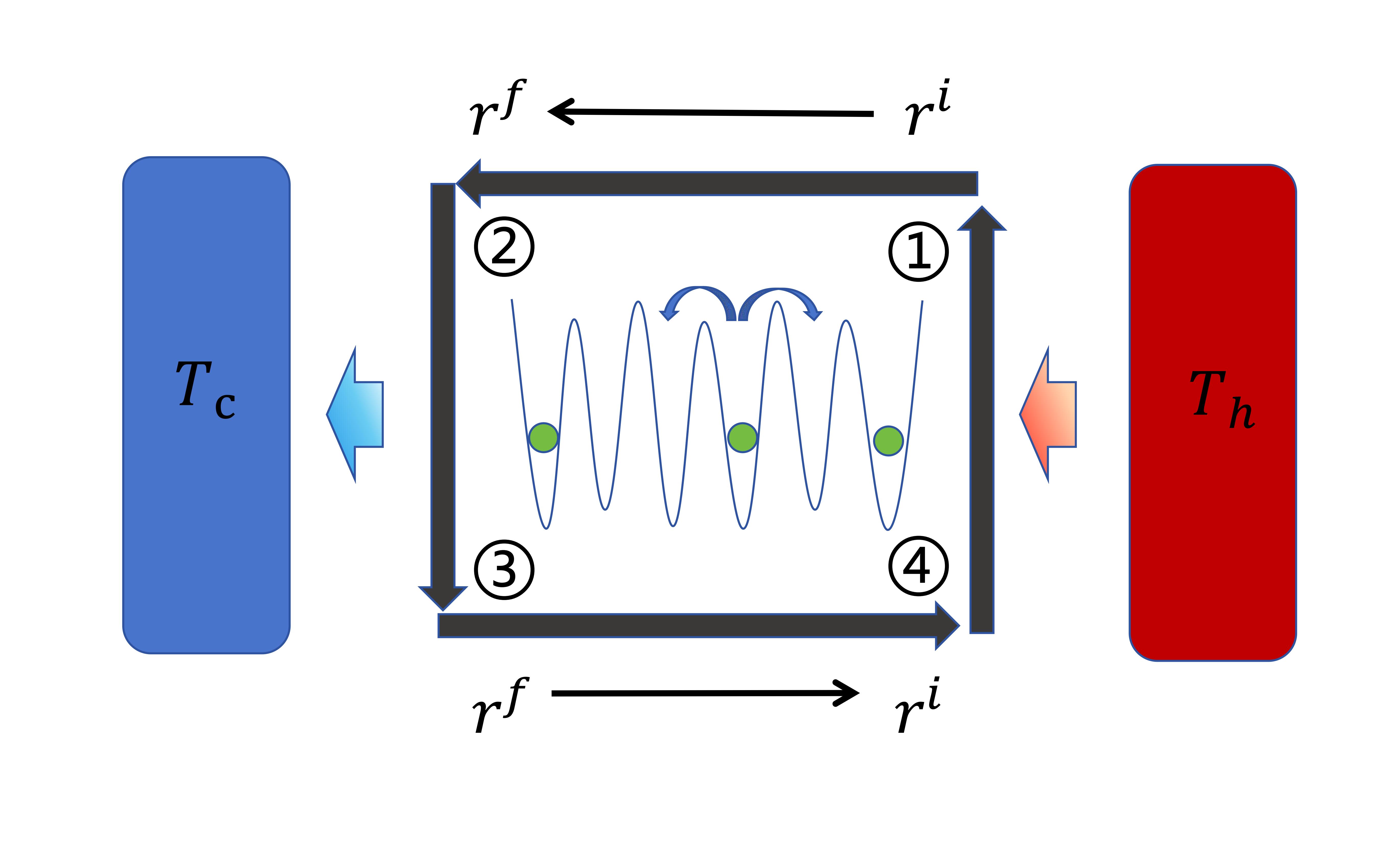}\\
		\caption{(Color Online) Schematic illustration of the four-stroke
quantum heat cycle process. $T_{h}$ and $T_{c}$ stand for the high-temperature
and low-temperature heat sources, respectively. The working medium employed
is the generalized AA model with diagonal and off-diagonal quasiperiodic
modulations. $r^{i}$ and $r^{f}$ represent the systematic parameters of the
corresponding Hamiltonians. The system size is $L=610$.
}
		\label{f3}
\end{figure}

We adopt this quasiperiodic system as the working medium to establish a
quantum heat cycle process. The schematic diagram of this cycle is illustrated
in Fig.~\ref{f3}, which can be considered the quantum analog \cite{ref52,ref55,ref56,ref60} of the
classical Otto cycle. The first (\textcircled{4} $\rightarrow$ \textcircled{1})
and third (\textcircled{2} $\rightarrow$ \textcircled{3}) strokes take place under
thermal contact with high-temperature ($T_{c}$) heat baths. These strokes proceed
in the absence of external driving forces or particle exchange. From a thermodynamic
perspective, the second (\textcircled{1} $\rightarrow$ \textcircled{2}) and
fourth (\textcircled{3} $\rightarrow$ \textcircled{4}) strokes are
adiabatic (i.e., under thermal isolation). However, they may not be strictly
adiabatic in the quantum-mechanical sense, as quantum transitions can arise
during the practical evolutionary process. After the working medium undergoes this
complete cycle, we can determine the heat absorbed
$Q_{h}$ ($Q_{h}=E_{1}-E_{4}$) from the $T_{h}$ source, the heat released
$Q_{c}$ ($Q_{c}=E_{3}-E_{2}$) to the $T_{c}$ source, and the net work $W$ done by the
working medium, where $W=Q_{h}+Q_{c}$. Here, $E_{1}$, $E_{2}$, $E_{3}$, and $E_{4}$ are
the energies of the four stages, i.e., \textcircled{1}, \textcircled{2}, \textcircled{3},
and \textcircled{4}, respectively. Notably, the heat cycle process complies with the
Clausius inequality, which serves as a fundamental cornerstone of thermodynamics.
Based on the specific values of $Q_{h}$, $Q_{c}$, and $W$, the engine that employs
the extended-localized (critical) quasiperiodic system as its working medium demonstrates
distinct operational modes \cite{ref59,ref60}:
\begin{enumerate}
  \item {\it Heat engine}: $Q_{h}>0$, $Q_{c}<0$, and $W>0$;
  \item {\it Refrigerator}: $Q_{h} <0$, $Q_{c}>0$, and $W<0$;
  \item {\it Heater}: $Q_{h}<0$, $Q_{c}<0$, and $W<0$;
  \item {\it Accelerator}: $Q_{h}>0$, $Q_{c}>0$, and $W<0$.
\end{enumerate}

In the first stroke, the working medium, which is characterized by the
Hamiltonian $H(r^{i})$, will ultimately attain thermal equilibrium and
relax into a Gibbs state. The density matrix $\rho_{1}$ of this state is
expressed as $\rho_{1}=\frac{e^{-\beta_{h}H(r^{i})}}{Z_{1}}$, where
$\beta_{h}=\frac{1}{k_{b}T_{h}}$ ($k_{b}$ denotes the Boltzmann constant)
and $Z_{1}={\rm Tr}\left[e^{-\beta_{h}H(r^{i})}\right]$ represents the partition
function. Therefore, at thermal equilibrium, the system energy is given
by $E_{1}={\rm Tr}\left[\rho_{1}H(r^{i})\right]$. In the third stroke,
the medium with $r=r^{f}$ is brought into contact with a heat source
featuring $\beta_{c}=\frac{1}{k_{b}T_{c}}$. Consequently, the corresponding
Gibbs state is $\rho_{3}=\frac{e^{-\beta_{c}H(r^{f})}}{Z_{3}}$, with the
partition function defined as $Z_{3}={\rm Tr}\left[e^{-\beta_{c}H(r^{f})}\right]$.
The energy of the medium at this stage is then $E_{3}={\rm Tr}\left[\rho_{3}H(r^{f})\right]$.

For the second and fourth strokes, we first consider the near-adiabatic scenario,
where the density matrix remains approximately unchanged throughout the evolutionary
process. This idea can be realized in the future by making use of the continuously developing
quantum feedback technology \cite{sayrinRealtimeQuantumFeedback2011,vijayStabilizingRabiOscillations2012,horowitzQuantumEffectsImprove2014,
zhangQuantumFeedbackTheory2017,uysQuantumControlMeasurement2018,
prechQuantumThermodynamicsContinuous2025}.
During the second stroke, the Hamiltonian parameter
is switched from $r^{i}$ to $r^{f}$. In this stroke, only work is performed without any
heat exchange occurring. The Gibbs state $\rho_{2}$ in this stroke remains approximately
unaltered, i.e., $\rho_{2}\approx\rho_{1}$, while the energy $E_{2}$ is updated to
$E_{2}={\rm Tr}\left[\rho_{2}H(r^{f})\right]$. The fourth stroke can be interpreted as a
thermal annealing process. In this process, the hopping parameter of the medium is switched
back from $r^{f}$ to $r^{i}$; however, the Gibbs state $\rho_{4}$ remains approximately equal
to $\rho_{3}$ ($\rho_{4}\approx\rho_{3}$). Thus, the energy of the medium
becomes $E_{4}={\rm Tr}\left[\rho_{4}H(r^{i})\right]$.

\begin{figure}[htp]
		\centering
		\includegraphics[width=0.5\textwidth]{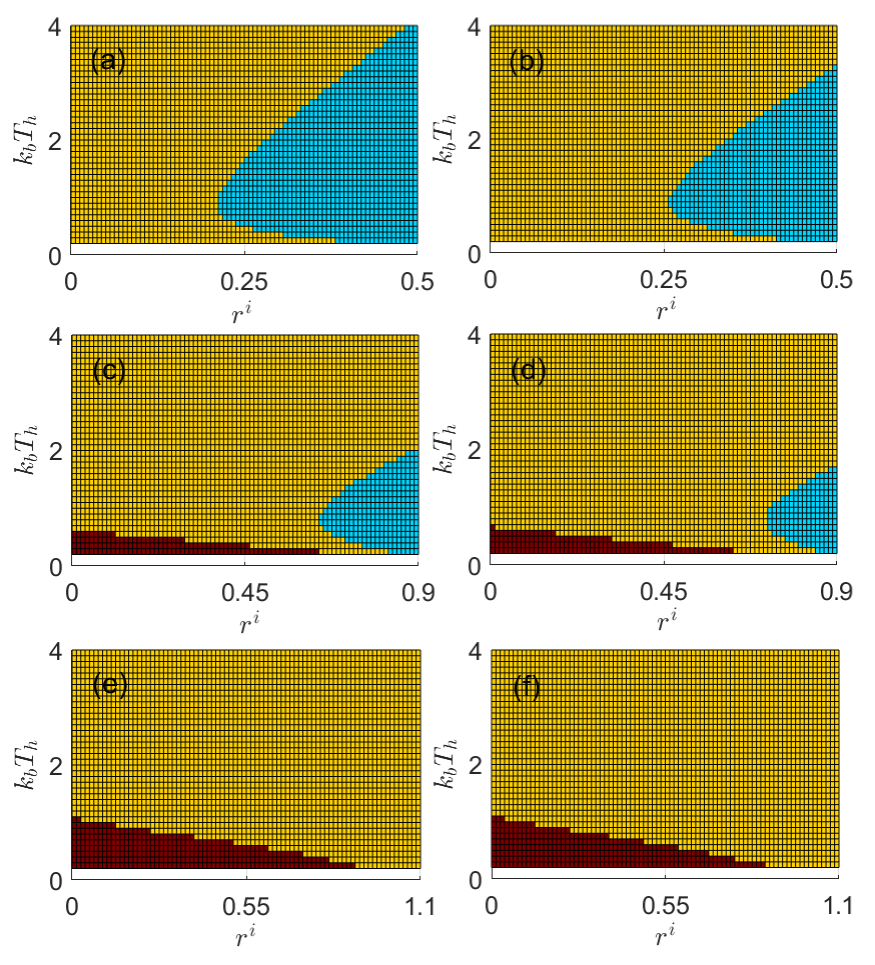}\\
		\caption{(Color Online) Working modes of the four-stroke cycle
in the near-adiabatic case.
(a) $r^{f}=0.5t$ and $\theta=\pi/6$. (b) $r^{f}=0.5t$ and $\theta=\pi/4$.
(c) $r^{f}=0.9t$ and $\theta=\pi/6$. (d) $r^{f}=0.9t$ and $\theta=\pi/4$.
(e) $r^{f}=1.1t$ and $\theta=\pi/6$. (f) $r^{f}=1.1t$ and $\theta=\pi/4$.
The blue regions represent the {\it Heat engine}. The yellow regions
denote the {\it Accelerator} and the brown regions denote the {\it Heater}.
Other parameters are $L=610$ and $k_{b}T_{c}=0.1t$.
}\label{f4}
\end{figure}

For a system with size $L=610$ and $k_{b}T_{c}=0.1t$, after
analyzing the values of $Q_{h}$, $Q_{c}$ and $W$, the
corresponding working modes of the four-stroke cycle under different
$r^{f}$ and $\theta$ are plotted
in Figs.~\ref{f4}(a)-\ref{f4}(f), respectively. It can be observed
that this cycle exhibits a diverse range of working modes. The blue
regions represent the {\it Heat engine} mode. The yellow regions
correspond to the {\it Accelerator} mode, while the brown
regions indicate the {\it Heater} working mode. It is noted from
Figs.~\ref{f4}(a)-\ref{f4}(d) that when $r^{f}$ and $r^{i}$ are
below the critical value $r_{c}$, the {\it Heat engine} mode appears.
This suggests that similar to the results of the critical-extended
quasi-periodic system \cite{ref40_1}, for the quasi-periodic system under
a finite $V$, its extended phase is favorable for the design of a
quantum heat engine as well. As $r^{f}$ increases and surpasses the
critical values $r_{c}=t$, the system enters the localized phase.
From Figs.~\ref{f4}(e) and \ref{f4}(f), we can see that regions
representing the {\it Heater} mode expand. This indicates that the localized
phase is more conducive to the realization of a quantum heater.
Additionally, it is clear that there are extensive parameter ranges
corresponding to the {\it Accelerator} mode, regardless of whether
the system is in the extended phase or the localized phase. Combining
with previous research on extended-critical quasiperiodic system \cite{ref40_1},
we draw a conclusion that all the extended, critical and localized
phases facilitate the realization of a quantum accelerator, and both
the localized and critical phases are conducive to the realization of
quantum heater.

Moreover, the results shown in Fig.~\ref{f4} offer strategies for
regulating the working mode of the four-stroke cycle. The transition
between different working modes can be achieved by tuning $r^{f}$,
$r^{i}$, and $k_{b}T_{h}$. For example, when $r^{f}$ is much smaller
than $r_{c}$ ( refer to Figs.~\ref{f4}(a) and \ref{f4}(b)), or vice
versa. When $r^{f}$ is close to but still less than $r^{c}$ and $k_{b}T_{h}$
is small (refer to Figs.~\ref{f4}(c) and \ref{f4}(d)), there are three
working modes. Thus, the cycle's working mode will change from {\it Heater}
to {\it Accelerator} and then to {\it Heat engine} when we gradually
increase $r^{i}$. When $r^{f}$ is larger than $r_{c}$, the cycle can be
toggled between {\it Heater} and {\it Accelerator} modes by tuning $r^{i}$
or $k_{b}T_{h}$.

\begin{figure}[htp]
		\centering
		\includegraphics[width=0.5\textwidth]{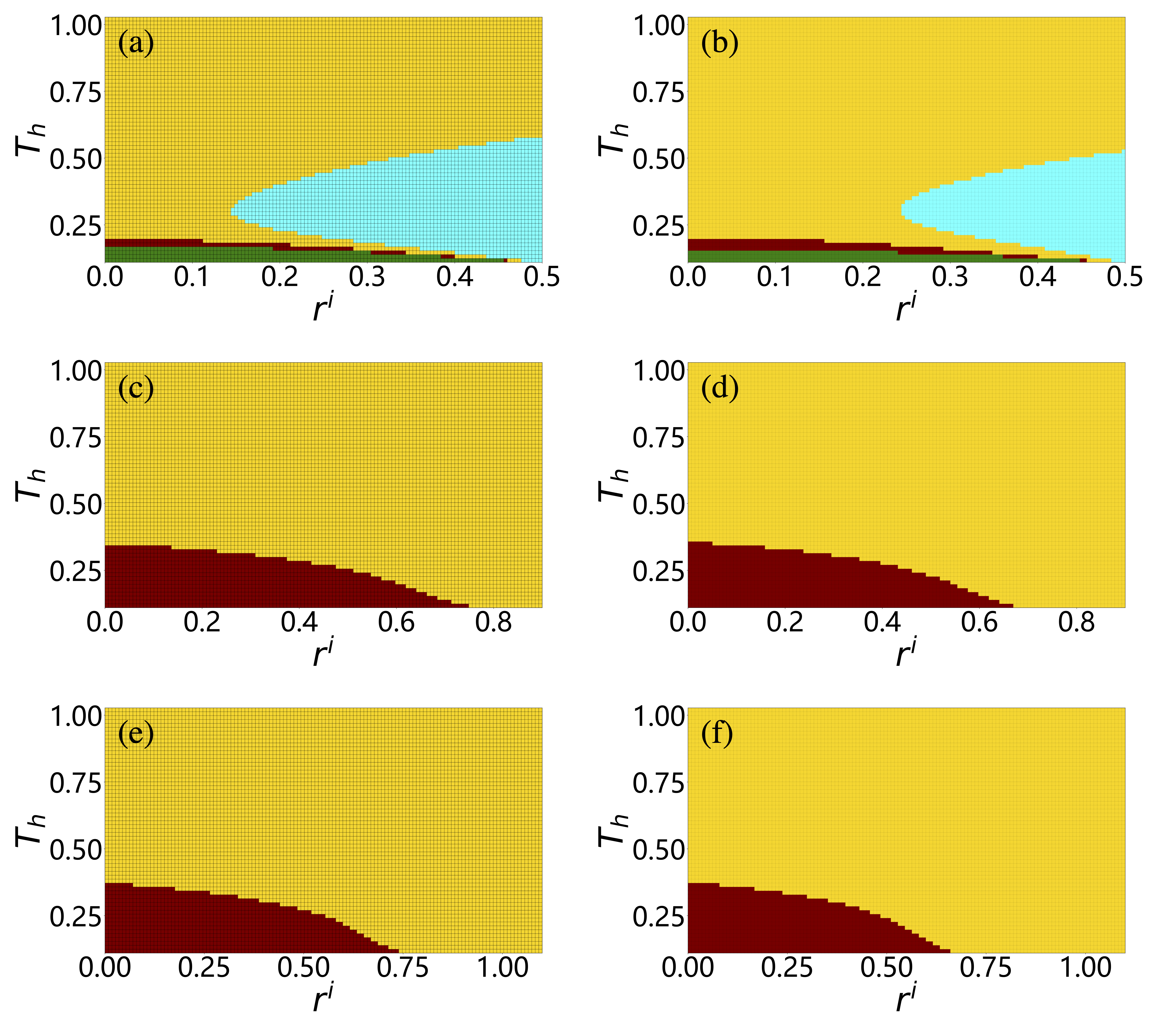}\\
		\caption{(Color Online) Working modes of the four-stroke cycle
in the adiabatic case.
(a) $r^{f}=0.5t$ and $\theta=\pi/6$. (b) $r^{f}=0.5t$ and $\theta=\pi/4$.
(c) $r^{f}=0.9t$ and $\theta=\pi/6$. (d) $r^{f}=0.9t$ and $\theta=\pi/4$.
(e) $r^{f}=1.1t$ and $\theta=\pi/6$. (f) $r^{f}=1.1t$ and $\theta=\pi/4$.
The blue regions represent the {\it Heat engine}. The green regions represent the {\it Refrigerator}.
The yellow regions denote the {\it Accelerator} and the brown regions denote the {\it Heater}.
Other parameters are $L=610$ and $k_{b}T_{c}=0.1t$.
}\label{f6}
\end{figure}

Next, we study the adiabatic process. The calculation steps for energy $E_{1}$ in the
first stroke and energy $E_{3}$ in the third stroke are consistent with those in the
near-adiabatic case. In the second stroke, we assume that
the system evolves from the $r^{i}$ state to the $r^{f}$ state in the quantum adiabatic form.
Therefore, the occupation probability of the particles remains unchanged.
By using the method of statistical mechanics, the partition function $Z_{2}$ is
derived as $Z_{2}=\prod^{L}_{j=1}\left[e^{-\beta_{2} E^{i}_{j}}\right]$, where $\beta_{2}=\frac{1}{k_{b}T_{h}}$,
and $E^{i}_{j}$ denotes $j$-th energy level of the working medium in the $r^{i}$ state.
Further, the energy $E_{2}$ is obtained as
$E_{2}=\sum^{L}_{j=1}E^{f}_{j}f(E^{i}_{j})$, where $E^{f}_{j}$ denotes the
$j$-th energy level of the working medium in the $r^{f}$ state and $f(E^{i}_{j})=\frac{1}{1+e^{\beta_{2} E^{i}_{j}}}$
represents the Fermi-Dirac distribution function.
Similarly, in the fourth stroke, we assume that the system also evolves
from the $r^{f}$ state to the $r^{i}$ state in the form of quantum adiabatic.
Thus, during the evolutionary process, the occupation probability of particles
at each energy level remains unchanged. By using the method of statistical physics,
the partition function $Z_{4}=\prod^{L}_{j=1}\left[e^{-\beta_{4} E^{f}_{j}}\right]$,
where $\beta_{4}=\frac{1}{k_{b}T_{c}}$, is obtained. Further, we obtain the energy
$E_{4}$ as $E_{4}=\sum^{L}_{j=1}E^{i}_{j}f(E^{f}_{j})$,
where $f(E^{f}_{j})=\frac{1}{1+e^{\beta_{4} E^{f}_{j}}}$ is the Fermi-Dirac distribution function.

Still taking system size $L=610$ and $k_{b}T_{c}=0.1t$, the working modes of four-stroke cycles
under different $r^{f}$and $\theta$ are shown in Figs.~\ref{f6}(a)-\ref{f6}(d).
It is readily seen that similar to the near-adiabatic case,the {\it Accelerator} mode
occupies the vast majority of the parameter space and is almost independent of the
selection of $r^{f}$ and $\theta$, indicating that the {\ it Accelerator} mode does not
depend on the properties of the system. The left working modes show obvious parameter
dependencies. It can be seen from Figs.~\ref{f6}(a) and \ref{f6}(b) that the {\it Heat engine}
mode only exists in the extended phase regime. However, compared with the near-adiabatic case,
the {\it Heat engine} mode in the adiabatic case is more sensitive to changes in the
parameter $r^{f}$. As can be seen from Figs.~\ref{f6}(a)-\ref{f6}(d), when $r^{f}$
is slightly increased, the {\it heat engine} mode disappears. Moreover, it can be seen
that the adiabatic condition compresses the high-temperature threshold of the {\it Heat engine} mode,
making the {\it Heat engine} in the adiabatic condition only operate at a lower
heat source temperature.
As can be seen from Figs.~\ref{f6}(e) and \ref{f6}(f), as the working medium in contact
with the low-temperature heat source enters the scope of the localized phase,
the working area of the {\it Heater} mode increases, indicating that the localized phase
is conducive to the existence of the {\it Heater} mode. However, compared with the
near-adiabatic case, the working area slightly decreases. Surprisingly, after taking
into account the adiabatic process, the {\it Refrigerator} mode emerges in the extended phase regime,
as shown in the green areas in Figs.~\ref{f6}(a) and \ref{f6}(b). However, the {\it Refrigerator}
is also parameter-sensitive. As $r^{f}$ increases, the {\it Refrigerator} mode disappears. Moreover,
a higher heat source temperature is also not conducive to the emergence of the {\it Refrigerator}.
But at least, these findings indicates that the adiabatic evolution process and appropriate
parameter settings can trigger the {\it Refrigerator} mode.

\section{Summary}\label{S5}
In conclusion, we have investigated the localization and thermodynamic properties of a
quasicrystal system featuring simultaneous diagonal and off-diagonal quasiperiodic modulations.
We demonstrate that extended, critical, and localized phases
can be distinguished not only through the fractal dimension of wave functions but also via quantum phase-space methods,
including the Wigner distribution and Wigner entropy. Notably, the Wigner entropy is largest in the critical phase,
intermediate in the extended phase, and smallest in the localized phase, offering new insights into localization phenomena
in quasiperiodic and disordered systems. Employing this quasicrystal as the working medium in a quantum Otto cycle,
we confirm that the extended phase enables a quantum heat engine mode, while the localized phase supports a heater mode.
Surprisingly, through introducing the adiabatic process, we uncover a fourth operational mode in this system:
a quantum refrigerator mode. These results will contribute to the development and utilization
of quasicrystals in thermodynamics.

This research is supported by Zhejiang Provincial Natural Science Foundation of China under Grant No. LQN25A040012,
the start-up fund from Xingzhi College, Zhejiang Normal University, and the  National Natural Science Foundation of China under Grant No. 12174346.

\bibliography{references}

\end{document}